\documentclass{ws-procs975x65}

\begin{document}

\wstoc{Classical and Quantum Aspects of the Inhomogeneous Mixmaster Chaoticity}{R. Benini,  G. Montani}

\title{Classical and Quantum Aspects of the Inhomogeneous Mixmaster Chaoticity}

\author{R. Benini${^12\dag}$ and G. Montani${^23\ddag}$}

\address{$^1$Dipartimento di Fisica - Universit\`a di Bologna and INFN\\ Sezione di Bologna,
via Irnerio 46, 40126 Bologna, Italy\\
$^2$ICRA---International Center for Relativistic Astrophysics  
c/o Dipartimento di Fisica (G9) Universit\`a di Roma ``La Sapienza'',
Piazza A.Moro 5 00185 Roma, Italy\\
$^3$ENEA C.R. Frascati (U.T.S. Fusione), Via Enrico Fermi 45, 00044 Frascati, Roma, Italy
$^\dag$\email{riccardo.benini@icra.it}
$^\ddag$\email{montani@icra.it}}			

\begin{abstract}
We refine Misner's analysis of the classical and quantum Mixmaster in the fully inhomogeneous picture; we both connect the quantum behavior to
the ensemble representation, both describe the precise effect of the
boundary conditions on the structure of the quantum states.
\end{abstract}

\bodymatter

\begin{figure}[ht]
   \begin{minipage}[b]{0.5\textwidth}
   \begin{flushleft}
	\includegraphics[width=0.8\hsize]{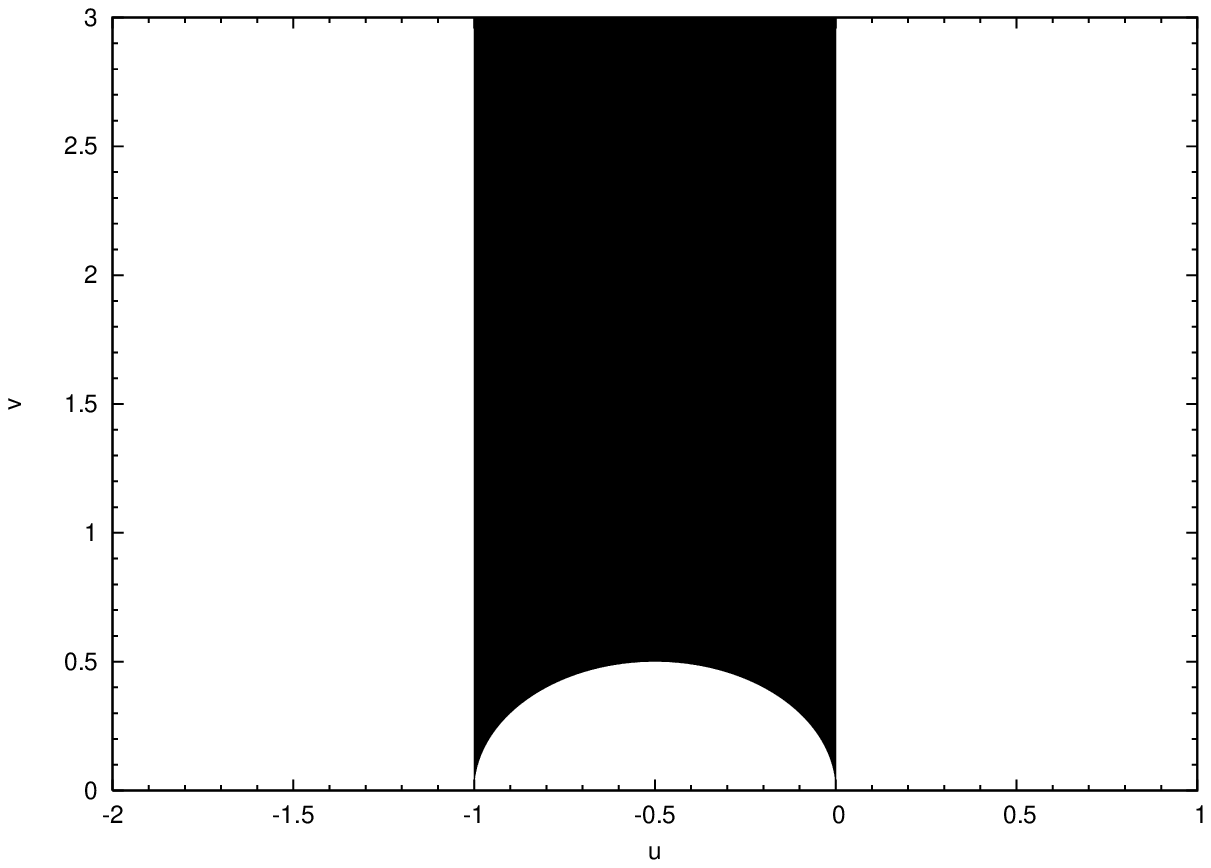} 
	\caption{The real domain $\Gamma_Q$\label{pot1.eps}}
	\end{flushleft}
	\end{minipage}%
	\begin{minipage}[b]{0.5\textwidth}
	\begin{flushright}
\includegraphics[width=0.8\hsize]{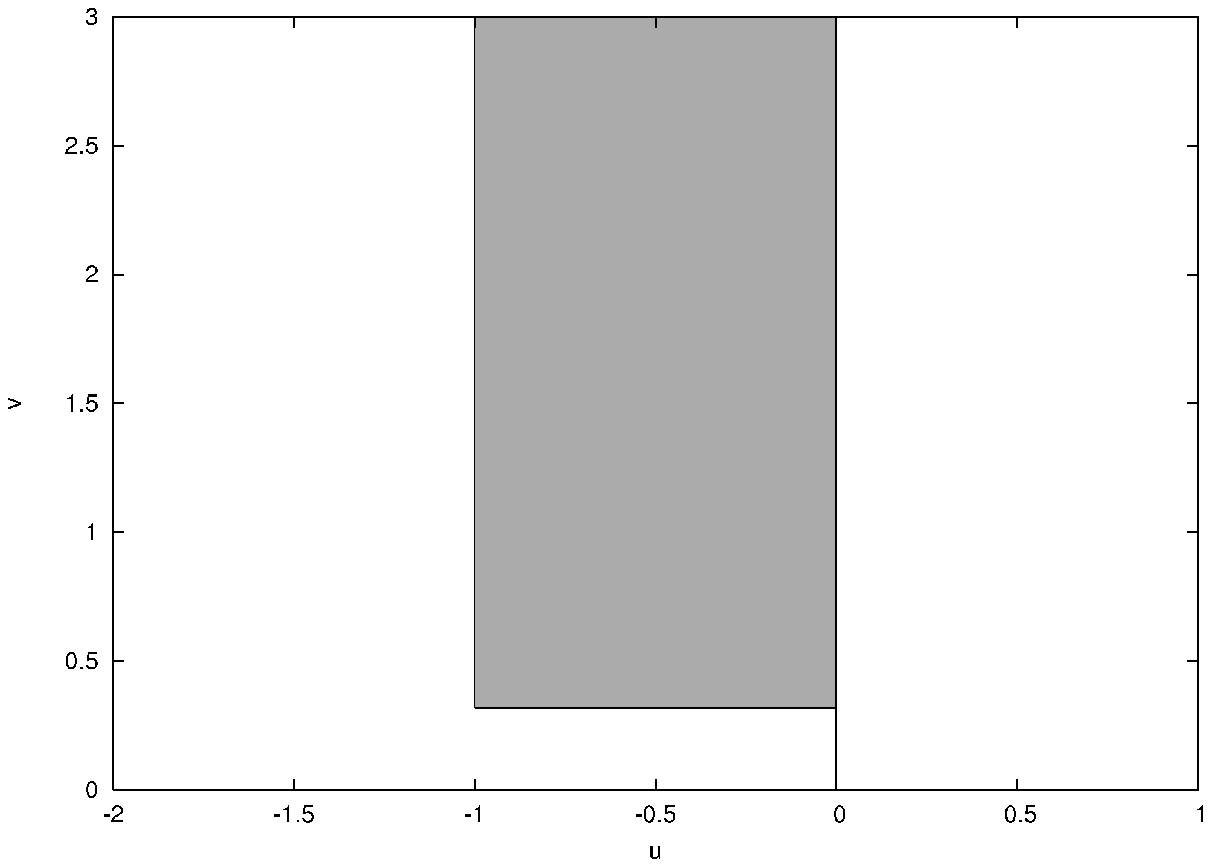} 
   \caption{The approximate domain\label{2}}
   \end{flushright}
   \end{minipage}
\end{figure}

Near the cosmological singularity, the dynamics of a generic inhomogeneous cosmological model is reduced by an ADM procedure to the evolution of the physical degrees of freedom, {\it i.e.} the anisotropies of the Universe. In fact asymptotically to the Big-Bang the space points dynamically decouple \cite{BeniniMontani2004PRD} because the spatial gradients of the dynamical variables become of higher order \cite{Kirillov1993ZETF} and we can model the inhomogeneous Mixmaster via the reduced action:
\begin{equation}
	\label{hamiltoniana uv}
I=\int d^3 x d\tau\left(p_u \partial_\tau u+p_v \partial_\tau v-\epsilon\right),\hspace{0.5cm}\epsilon=v \sqrt{p_u^2+p_v^2}\,,
\end{equation}
The dynamics of such a model is equivalent to (the one of) a billiard-ball on a Lobachevsky plane; this can be shown by the use of the Jacobi metric \cite{BeniniMontani2004PRD}. 
The manifold described turns out to have a constant negative curvature, where the Ricci scalar is given by $R=-2/E^2$: the complex dynamics of the generic inhomogeneous model results in a collection of decoupled dynamical systems, one for each point of the space, and all of them equivalent to a billiard problem on a Lobachevsky plane.\\
We want to investigate the relation existing between the classical and the semiclassical dynamics, and from this analysis we will derive the correct operator ordering to be used when quantizing the system.\\
Let's write down the Hamilton-Jacobi equation for the system
\begin{equation}
\label{equazione per S}
\epsilon^2 = v^2 \left(\left({\frac{\delta {\mathcal S}_0}{\delta u}}\right)^2+\left({\frac{\delta {\mathcal S}_0}{\delta v}}\right)^2\right)
\end{equation}
This can be explicitly solved by separation of constants
\begin{equation}
\label{S}
{\mathcal S}_0(u,v)=k(y^a) u + \sqrt{\epsilon^2-k^2(y^a) v^2} -\epsilon \ln \left(2\frac{\epsilon+\sqrt{\epsilon^2-k^2(y^a) v^2}}{\epsilon^2 v}\right)+c(y^a)\,.
\end{equation}
The semiclassical analysis can be developed furthermore, and the stationary continuity equation for the distribution function can be worked in order to obtain informations about the statistical properties of the model; as soon as we restrict the dynamics to the configuration space, we get the following equation for the distribution function $\tilde{w}$
\begin{equation}
\label{eq w}
\displaystyle\frac{\partial \tilde{w}(u,v;k)}{\partial u}+\sqrt{\left(\frac{E}{k v}\right)^2-1}\displaystyle\frac{\partial \tilde{w}(u,v;k)}{\partial v}+\frac{E^2-2 k^2v^2}{k v^2} \frac{\tilde{w}(u,v;k)}{\sqrt{E^2-(k v)^2}}=0
\end{equation}
This can be solved, and the exact distribution function can be obtained as soon as we eliminate by integration the constant $k$
\begin{equation}
	\label{soluzione eq w}
	\tilde{w}(u,v)=\int_{-\frac{E}{v}}^{\frac{E}{v}}\frac{g\left(u+v\sqrt{\frac{E^2}{k^2 v^2}-1}\right)}{v \sqrt{E^2-k^2 v^2}}dk
\end{equation}
It is worth nothing how in the case $g=const$, the microcanonical Liouville measure $w_{mc}(u,v)=\frac{\pi}{v^2}$ is recovered.\\
We expect that the distribution function $\tilde{w}(u,v)$ is re-obtained as soon as the quantum dynamics is investigated to the first order in $\hbar$. 
This can be easily done as a WKB approximation to the quantum dynamics is constructed; as soon as we retain only the lower order in $\hbar$, we obtain that: i) the phase  $S(u,v)$ coincides with the Hamilton-Jacobi function, and ii) the probability density function $r(u,v)$ obeys the following equation:
\begin{equation}
\label{eq per r}
k \displaystyle\frac{\partial r(u,v)}{\partial u}+\sqrt{\left(\frac{E}{v}\right)^2-k^2} \displaystyle\frac{\partial r(u,v)}{\partial v}+\frac{a(E^2-k^2 v^2)-E^2}{v^2 \sqrt{E^2-k^2 v^2}} r(u,v)=0\,.
\end{equation}
that coincides with (\ref{eq w}) for a particular choice for the operator-ordering only, {\it i.e.}
\begin{equation}
	\label{ordinamento operatoriale}
	\hat{v}^2 \hat{p}_v^2\,\rightarrow\,-\hbar^2 \frac{\partial}{\partial v}\left( v^2\frac{\partial}{\partial v}\right)\,.
\end{equation}
With this result, the problem of the full quantization of the system can be taken in consideration. The main problem is the presence of the root square function in the definition of the Hamiltonian 	(\ref{hamiltoniana uv}), but well grounded motivations exist\cite{Puzio1994CQG} to assume that the real Hamiltonian and the squared one  have same eigenfunctions and squared eigenvalues. This way the solution of the eigenvalue equation can be obtained
\begin{equation}
	\label{relazione sulle k}
	\Psi(u,v)=\sum_{n>0}a_n K_{s-1/2}(2n \pi v) \sin (2n\pi u)
\end{equation}
The spectrum is obtained as Dirichelet boundary conditions on the domain $\Gamma_Q$ (see Fig.\ref{pot1.eps}) are taken into account. The condition on the vertical lines can be imposed exactly, but the one on the semicircle cannot be solved exactly; so we approximate it as in Fig.\ref{2} with a straight line $v=1/\pi$   (it preserves the domain area $\mu=\pi$).\\
All these imply that $s=1/2+t$, with $t\in\Re$, and that the spectrum assumes the following form
\begin{equation}
	\label{autovalori}
	E_t^2=\displaystyle(\frac{1}{4}+t^2)\hbar^2
\end{equation}
The values of the real parameter $t$ have to be numerically evaluated by solving the equation $K_{it}(2 n)=0$ for every natural $n$. This condition implies a discrete but quite complicated shape for the spectrum.\\
Asymptotic expansions for high occupation numbers can be derived for different regions of the parameters $(t,\;n)$\cite{BeniniMontani2006CQG}

\end{document}